\documentclass[twocolumn,10pt]{IEEEtran}
\renewcommand{\baselinestretch}{0.99}
\usepackage{latexsym}
\usepackage{amsmath, graphics,amssymb, subfig}
\usepackage{cite}
\usepackage{url}
\usepackage{color,soul}
\usepackage{hhline}
\usepackage{bbold}
\usepackage{algorithm}
\usepackage{algorithmic}
\usepackage{graphicx}
\usepackage{bm}
\usepackage{float}
\usepackage{overpic}

\def\BibTeX{{\rm B\kern-.05em{\sc i\kern-.025em b}\kern-.08em
    T\kern-.1667em\lower.7ex\hbox{E}\kern-.125emX}}

\newfont{\bb}{msbm10 scaled 1000}

\newcommand{\be}{\begin{equation}}
\newcommand{\ee}{\end{equation}}
\newcommand{\bea}{\begin{eqnarray}}
\newcommand{\eea}{\end{eqnarray}}
\newcommand{\bitem}{\begin{itemize}}
\newcommand{\eitem}{\end{itemize}}

\makeatletter
\def\hlinewd#1{%
\noalign{\ifnum0=`}\fi\hrule \@height #1 \futurelet \reserved@a\@xhline}

\makeatother

\makeatletter

\newcommand*{\rom}[1]{\expandafter\@slowromancap\romannumeral #1@}
\makeatother

\usepackage{multirow}
\usepackage{diagbox}
\usepackage{makecell}

\begin{document}

\setlength{\abovedisplayskip}{5pt}
\setlength{\belowdisplayskip}{5pt}

\title{{\LARGE Deep Reinforcement Learning for Adaptive Power Allocation \\ in ISAC Systems with Mobile Target}}
\vspace{0mm}

\author{\IEEEauthorblockN{Zhilin Fu, Sangmin Kim, Sangwon Hwang, \textit{Member}, \textit{IEEE}, Jihwan Moon, \textit{Member}, \textit{IEEE}, Jeongwon Kim, Jaewan Kim, and Inkyu Lee, \textit{Fellow}, \textit{IEEE} \\ \vspace*{-6mm}}
% \thanks{Manuscript received Month Date, 20xx; revised Month Date, 20xx;
% accepted Month Date, 20xx. Date of publication Month Date, 20xx;
% date of current version Month Date, 20xx. This work was supported in part
% by XXX. The associate editor coordinating
% the review of this article and approving it for publication was XXX.
% (Corresponding author: Inkyu Lee.)}
\thanks{This work was supported in part by the National Research Foundation of Korea (NRF) Grant funded by the Korea Government (MSIT) under Grant RS2022-NR070834, and in part by the Institute of Information \& Communications Technology Planning \& Evaluation (IITP)–Information Technology Research Center (ITRC) Grant funded by the Korea Government (MSIT) (IITP-2026-RS-2024-00437886, 50\%).}

\thanks{Z. Fu is with the School of Electrical Engineering, Korea University, Seoul 02841, South Korea, and with Samsung Electronics, Suwon 16677, South Korea  (e-mail: celynnfu@korea.ac.kr).}

\thanks{S. Kim, Jeongwon Kim and I. Lee are with the School of Electrical Engineering, Korea University, Seoul 02841, South Korea (e-mail: \{smgeem; jeongwonkim; inkyu\}@korea.ac.kr).}

\thanks{S. Hwang is with the Department of Computer and Artificial Intelligence Engineering, Pukyong National University, Busan, 48513, South Korea (e-mail: s.won.hwang@pknu.ac.kr).}

\thanks{J. Moon is with the Department of Mobile Convergence Engineering, Hanbat National University, Daejeon 34158, South Korea (e-mail: \nobreak anschino@staff.hanbat.ac.kr).}

\thanks{Jaewan Kim is with the Augmented Cognition Meta-Communications ERC Research Center, Korea University, Seoul 02841, South Korea (e-mail: \nobreak kupjohnkim@korea.ac.kr).}

\thanks{(Corresponding authors: Inkyu Lee.)} 

\thanks{Copyright (c) 20xx IEEE.}
}\maketitle

%%%%%%%%%%%%%%%%%%%%%%%%%%%%%%%%%%%%%%%%% Roman Number
\newcommand{\RNum}[1]{\uppercase\expandafter{\romannumeral #1\relax}}
%%%%%%%%%%%%%%%%%%%%%%%%%%%%%%%%%%%%%%%%%%%%%%%%%%%%%%%%%%%%%%%%%%%%%%%%%%%%%%%%%%%%%%%%%%%%%%%%%%%%%%%ABStract
\begin{abstract}
In this paper, we study the power allocation for an integrated sensing and communication (ISAC) system which tracks a mobile target. We first model the problem as a Markov decision process, and then tackle it with a soft actor-critic (SAC) based deep reinforcement learning (DRL) approach. We also combine a Dirichlet policy, which naturally produces normalized continuous actions under random target motion. To exploit different features of sensing and communication operations, we carefully design a reward function such that the system can dynamically control power allocation to conserve resources. The simulation results demonstrate that the proposed scheme enhances tracking performance compared to other baselines while sustaining communication performance.
\vspace*{-1mm}
\end{abstract}

% \begin{IEEEkeywords}
% Integrated sensing and communication, resource allocation, deep reinforcement learning.
% \end{IEEEkeywords}

\vspace{-4mm}
\section{Introduction} 
\label{sec:introduction}
\vspace{-1mm}

Integrated sensing and communication (ISAC) has been recognized as a key technology for future wireless systems, aiming to unify sensing and communication within a single framework \cite{JZhang:22}. By sharing spectrum and hardware, the ISAC can improve both spectral and energy efficiency, and support various applications such as autonomous driving, smart homes, and industrial automation, with improved service quality and reduced infrastructure costs. Its feasibility has been discussed in many studies, either by extending existing communication infrastructures with sensing capabilities \cite{KAbboud:16, MRahman:20TAES, PKumari:18TVT} or by pursuing deeper integration in future network architectures \cite{PRaviteja:18TWC, NGP:24Proc, JLi:24IOTJ, CBarneto:22TCom}.

In practice, sensing for target tracking is carried out over a long period of time, whereas communication slots last from microseconds to milliseconds with high quality of service (QoS) requirements \cite{ZXiao:25TWC, AGraff:23TVT, KMeng:23TWC}. This timescale difference can induce redundant sensing and waste energy if both operations run in every slots \cite{KMeng:23TWC, LabSubin:21TVT}. For example, in typical communication systems, the slot duration is on the order of milliseconds \cite{SlotLength5G}. At this fine scheduling granularity, if the sensing is carried out at all time, it incurs unnecessary energy consumption, while an additional tracking gain would be marginal. Moreover, developing a solution becomes more challenging when the target is no longer static. As the target moves, prediction errors increase uncertainty, which makes conventional sensing fail to satisfy real-time tracking demands.

Previous studies have investigated ISAC resource allocation from two main directions: model-based optimization \cite{KMeng:23TWC, ZPu:24TGCN, FDong:23TWC, MLi:25WCL} and learning-based control \cite{JLee:22TVT, YWang:24TWC}. For model-based optimization, \cite{ZPu:24TGCN} optimized a trade-off between the communication rate and the sensing Fisher information (FI). In \cite{KMeng:23TWC}, an integrated periodic sensing and communication (IPSAC) scheme was introduced, where sensing and communication are activated on demand to reduce energy consumption, and the communication rate is maximized under a sensing signal-to-noise ratio (SNR) constraint for static targets. However, it does not address mobile target tracking, where the sensing requirement changes over time according to the target motion and the resulting tracking uncertainty. Many existing ISAC designs rely on instantaneous sensing metrics evaluated per time slot such as SNR and FI, which may not sufficiently capture temporal uncertainty evolution.

To address mobility, some recent works adopted the posterior Cramér–Rao bound (PCRB) \cite{FDong:23TWC, MLi:25WCL}, which accounts for both the motion dynamics and the prior estimation covariance to reflect the achievable tracking accuracy. In \cite{FDong:23TWC}, a prediction-driven framework was introduced that jointly manages resources using the predicted PCRB, while \cite{MLi:25WCL} employed the predicted PCRB for beamforming design in roadside unit based ISAC systems. However, due to the discrepancy between the predicted and actual PCRB, such model-based optimizations cannot accurately represent or improve the true tracking performance.

These limitations motivate learning-based approaches that can acquire effective sensing–communication strategies directly from interaction with the environment, without relying on instantaneous or predicted metrics. Reinforcement learning (RL) provides such flexibility by learning sequential decision policies in dynamic environments \cite{LabMintae:24IOT, LabSangwon:25IOT}. For example, the authors in \cite{JLee:22TVT} developed a deep RL (DRL) framework with graph neural networks for autonomous vehicle tracking, introducing a data usefulness metric that incorporates both age of information and spatial signatures. However, such metrics mainly reflect data freshness and spatial features, rather than a dedicated tracking-reliability metric. In \cite{YWang:24TWC}, a hybrid convex–RL method was proposed for resource allocation in bi-static ISAC with mobile target tracking, but its use of dynamic PCRB thresholds and prediction-based RL states may lead to tracking instability and decision uncertainty.

In this paper, we propose an RL-based approach to adaptively balance sensing and communication while tracking mobile targets. The main contributions are as follows:
\begin{itemize}
    \item We introduce a DRL-based power allocation algorithm for ISAC in mobile target tracking scenarios. The proposed scheme is trained offline to learn the correlations across time slots and applied online to improve resource efficiency without degrading communication reliability.
    \item Our reward design reflects the difference between sensing and communication requirements, allowing the system to adjust the sensing period when prior sensing information is sufficient and focus more resources on communication if necessary. The experimental results demonstrate that the proposed scheme is able to satisfy the actual PCRB constraint compared to conventional methods.
\end{itemize}
We define $\mathrm{diag} \left( \cdot \right)$ as a diagonal matrix containing each element on the diagonal. $\mathbf{A}^T$, $\mathbf{A}^H$ and $\mathrm{tr} \left( \mathbf{A} \right)$ represent transpose, Hermitian transpose, and trace, respectively. $\otimes $ denotes the Kronecker product. $\mathbb{E} \left[ \cdot \right] $ defines the expectation operation.

\vspace{-2mm}
\section{System Model} 
\label{sec:SystemMode}
\vspace{-1mm}

\begin{figure}
    \begin{center}
    \includegraphics[width=3.4in, trim=0 300 10 300, clip]{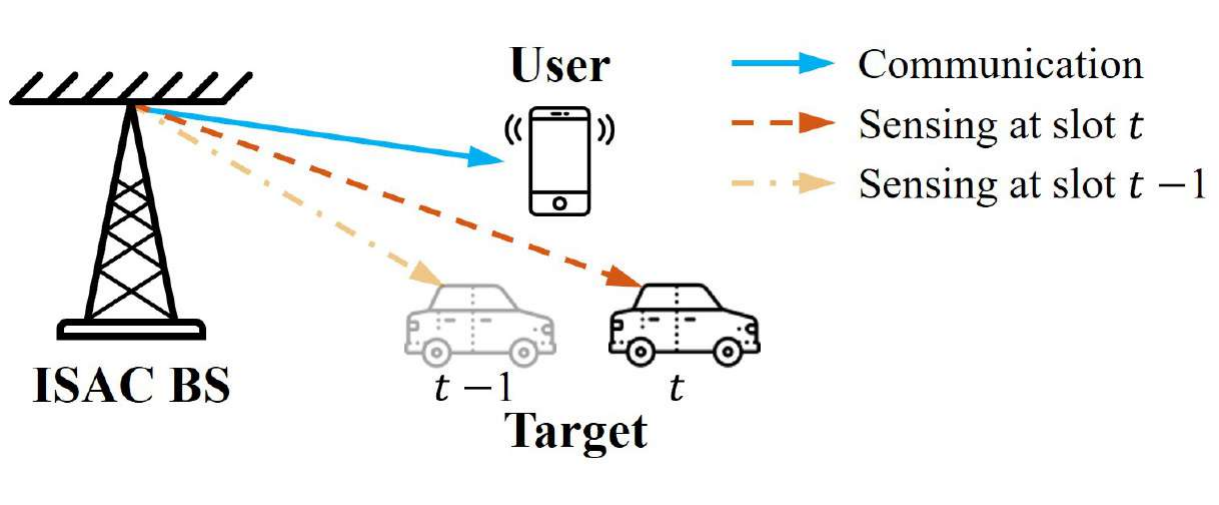} 
    \end{center}
    \vspace{-4mm}
    \caption{ISAC system model}
    \label{fig:SysMod}
    \vspace{-4mm}
\end{figure}

As shown in Fig. \ref{fig:SysMod}, we consider an ISAC system where a base station (BS) is equipped with $M$ antennas in a uniform linear array. The BS provides communication services to a single-antenna user while tracking a mobile target at the same time. The transmit and receive antenna steering vectors are given by 
\begin{equation}
\begin{aligned}
    \mathbf{a}_x \! \left( \theta \right)  = \! \frac{1}{\sqrt{M}} 
     \left[ 
     1, e^{j \frac{2\pi l}{\lambda} \sin \theta}, ...,  e^{j \frac{2\pi l}{\lambda} (M-1) \sin \theta}
     \right]^T,
\end{aligned}
\label{eq:antenna_steering}
\end{equation}
where $x$ represents $x \in \left\{\mathrm{t}, \mathrm{r} \right\}$, and $l$, $\lambda$, $\theta$ equal the antenna space, the wavelength, and the angle of departure, respectively.

Let us define $\theta_{\mathrm{c}}$ and $\theta_{\mathrm{s},n}$ as the actual angle of the user and the target at the $n$-th slot, respectively, and $\hat{\theta}_{\mathrm{c}}$ and $\hat{\theta}_{\mathrm{s}, n}$ as their estimated values. Then, the transmit beamforming vectors of communication and sensing can be represented as $\mathbf{u} = \mathbf{a}_{\mathrm{t}} ( \hat{\theta}_{\mathrm{c}} )$ and $\mathbf{w}_{\mathrm{t},n} = \mathbf{a}_{\mathrm{t}} ( \hat{\theta}_{\mathrm{s}, n} )$, respectively. The transmit signal of a BS at the $n$-th time slot is obtained by
\begin{equation}
    \mathbf{x}_n = \sqrt{p_{\mathrm{c}, n}} \mathbf{u} c_n + \sqrt{p_{\mathrm{s}, n}} \mathbf{w}_{\mathrm{t},n}  s_n,
    \label{eq:x_n}
\end{equation}
where $ p_{\mathrm{c}, n}$ and $ p_{\mathrm{s}, n}$ stand for the allocated power for communication and sensing, respectively, and $c_n$ and $s_n$ represent the transmitted signals for communication and sensing, respectively. Here, the total transmit power satisfies $p_{\mathrm{c}, n} + p_{\mathrm{s}, n} \leq P_{\mathrm{max}}$, where $P_{\mathrm{max}}$ is defined as the maximum transmit power.

In this paper, we assume that channel state information (CSI) of the user and the initial state of the target are perfectly known at the BS. The target state is characterized by its angle $\theta_{\mathrm{s}}$, the BS-target distance $d_{\mathrm{s}}$, and the target velocity $v_{\mathrm{s}}$. After the initial target-state acquisition, the system operates over $N$ time slots while continuously serving the UE. At each slot, the BS allocates power between communication and sensing according to the available tracking state. When sensing is activated, the BS operates in the ISAC mode and updates the target estimate from the received echo. Otherwise, it operates in the communication-only mode with $p_{\mathrm{s},n}=0$ and only predicts the target state using the motion model.

\vspace{-2mm}
\subsection{Communication Model}
\vspace{-1mm}

Denoting the path-loss factor as $ L \triangleq \left( \frac{4 \pi}{\lambda} \right)^{2} d_{\mathrm{c}}^{\eta}$ where $d_{\mathrm{c}}$ and $\eta$ equal the distance between the BS and the user, and the path-loss exponent, respectively, the channel vector $\mathbf{h}_{\mathrm{c}} \in \mathbb{C}^{M \times 1}$ is expressed by $\mathbf{h}_{\mathrm{c}} = \frac{\alpha_{\mathrm{c}}}{\sqrt{L}} \mathbf{a} \left( \theta_{\mathrm{c}} \right)$, where $\alpha_{\mathrm{c}}$ accounts for a random complex Gaussian coefficient with zero mean and unit variance. The received signal of the user at the $n$-th time slot is written by
\begin{displaymath}
\begin{aligned}
    y_{\mathrm{c}, n}
    &= \sqrt{p_{\mathrm{c}, n}} \mathbf{h}_{\mathrm{c}}^{H} \mathbf{u} c_n + \sqrt{p_{\mathrm{s}, n}} \mathbf{h}_{\mathrm{c}}^{H} \mathbf{w}_{\mathrm{t}, n} s_n + n_{\mathrm{c}, n},
\end{aligned}
\end{displaymath}
where $n_{\mathrm{c}, n} \sim \mathcal{CN} \left( 0 \right. ,\left. \sigma_{\mathrm{c}}^{2} \right)$ indicates the complex additive white Gaussian noise (AWGN). Finally, the achievable communication rate at the $n$-th time slot can be calculated as 
\begin{equation}
    C_n = \log_2 \left( 1 + \frac{ p_{\mathrm{c}, n} | \mathbf{h}_{\mathrm{c}}^{H} \mathbf{u}|^2 }{ p_{\mathrm{s}, n} |\mathbf{h}_{\mathrm{c}}^{H} \mathbf{w}_{\mathrm{t},n} |^2 + \sigma_{\mathrm{c}}^{2}} \right) .
\end{equation}

\vspace{-2mm}
\subsection{Sensing Model}
\vspace{-1mm}

The BS transmits the sensing beam toward the predicted target direction and then estimates the target's state from the received echo signal. We assume ideal downconversion and use the baseband equivalent of the received echo signal \cite{FDong:23TWC}, which removes the high-frequency carrier components. The resulting signal is given as
\begin{displaymath}
\begin{aligned}
    r_n (\!t\!) 
    =& \kappa \alpha_n e^{j2\pi t \nu_{n}} \! \sqrt{p_{\mathrm{s}, n}} \mathbf{w}_{\mathrm{r}, n}^{H} \mathbf{a}_{\mathrm{r}} \left( \! \theta_{\mathrm{s},n} \! \right)
    \mathbf{a}_{\mathrm{t}}^H \left( \! \theta_{\mathrm{s},n} \! \right) \mathbf{w}_{\mathrm{t}, n}  s_n ( \! t \!-\! \tau_{n} \!) \\
    &+ \mathbf{w}_{\mathrm{r}, n}^{H} n_{\mathrm{s}}(t) ,
\end{aligned}
\end{displaymath}
where $\kappa=M$, $\alpha_n$, $\nu_{n}$, $\mathbf{w}_{\mathrm{r}, n} = \mathbf{a}_{\mathrm{r}} ( \hat{\theta}_{\mathrm{s}, n} )$ and $\tau_{n}$ denote the array gain factor, the reflection coefficient, Doppler frequency, the received beamforming vector and the time delay of the target at the $n$-th slot, respectively, and $n_{\mathrm{s}}(t) \sim \mathcal{CN} \left( 0 \right. ,\left. \sigma_{\mathrm{s}}^{2} \right)$ is the complex AWGN \cite{FDong:23TWC}.  The reflection coefficient $\alpha_n$ is modeled as a zero mean complex Gaussian random variable with variance $\sigma_{\alpha_n}^2 = \beta_{n} / (2 d_{\mathrm{s}, n})$, where $\beta_{n}$ and $d_{\mathrm{s},n}$ represent the radar cross section and the distance between the BS and the target, respectively.  

Next, the baseband echo signal is processed with a spatial-temporal matched filter, which extracts the signal corresponding to a specific delay-Doppler bin $\left( \tau, \nu \right)$ \cite{FDong:23TWC}. The matched filter is designed to maximize the received power for each bin, and its SNR is written as \footnote{Here, $\tau$ and $\nu$ determine the delay-Doppler bin of the extracted target echo, and thus do not appear explicitly in the final SNR expression.}
\begin{equation}
    \mathrm{SNR}_n = \frac{
        p_{\mathrm{s}, n} \kappa^4 \gamma_{\mathrm{t}, n} \gamma_{\mathrm{r}, n} \sigma_{\alpha_n}^2
    }{ 
        M \sigma_{\mathrm{s}}^2
    } , 
    \label{eq:sens_ChannelGain}
\end{equation}
where $\gamma_{\mathrm{t}, n} \! = \! | \mathbf{a}_{\mathrm{t}}^H ( \! {\theta}_{\mathrm{s}, n} \! ) \mathbf{a}_{\mathrm{t}} ( \! \hat{\theta}_{\mathrm{s}, n} \! ) |^2$ and $\gamma_{\mathrm{r}, n} \! = \! | \mathbf{a}_{\mathrm{r}}^H ( \! {\theta}_{\mathrm{s}, n} \! ) \mathbf{a}_{\mathrm{r}} ( \hat{\theta}_{\mathrm{s}, n} ) |^2$ equal the beamforming gains for the transmitter and the receiver, respectively.

\vspace{-2mm}
\section{Target Tracking}
\label{sec:TargetTracking}
\vspace{-1mm}

In this section, we describe the target mobility model. Then, we apply the extended Kalman filter (EKF) for state estimation, which predicts with the motion model and updates from measurements using local linearization at the current estimate \cite{FDong:23TWC}. Let us denote the motion state vector of the target at the $n$-th slot as $\boldsymbol{\xi }_{n}=\left[ x_n, y_n, \dot{x}_n, \dot{y}_n \right] ^T$, where $(x_n, y_n)$ and $(\dot{x}_n, \dot{y}_n)$ indicate the position and velocity components in the Cartesian coordinate, respectively. Also, we define $g_n$ as the number of slots since the last sensing action, which means that the effective sensing interval is $g_n T_{\mathrm{s}}$. Denoting the index of the last slot with ISAC mode as $\hat{n}$, the target motion is given by
\vspace{-1mm}
\begin{equation}
    \boldsymbol{\xi }_{n} = \mathbf{F}_n \boldsymbol{\xi}_{\hat{n}} + \boldsymbol{\omega}_{\hat{n}} , 
    \label{eq:state_motion}
    \vspace{-1mm}
\end{equation}
where $\mathbf{F}_n$ defines the state transition matrix as
\begin{equation}
    \mathbf{F}_n = \left[ 
    \begin{matrix}
	1&		g_n T_{\mathrm{s}}\\
	0&		1\\
    \end{matrix} 
    \right] \otimes \mathbf{I}_2, 
\end{equation}
and $\boldsymbol{\omega}_{\hat{n}}$ equals the state noise following a zero-mean Gaussian distribution with the covariance 
\begin{equation}
    \mathbf{\Phi }_n = \left[ 
    \begin{matrix}
	{\frac{1}{3}} (g_n T_{\mathrm{s}})^{3}&		{\frac{1}{2}} (g_n T_{\mathrm{s}})^{2}\\
	{\frac{1}{2}} (g_n T_{\mathrm{s}})^{2}&		g_n T_{\mathrm{s}}\\
    \end{matrix} ,
    \right] \otimes q_{\mathrm{s}} \mathbf{I}_2
\end{equation}
with $q_{\mathrm{s}}$ being the process noise intensity. This state-transition model corresponds to a constant-velocity approximation with the process noise over the effective sensing interval $g_nT_{\mathrm{s}}$ \cite{FDong:23TWC}.

Based on the received echo signals, the range $d$, the radial velocity $v$, and the angle $\theta$ of the target can be estimated. According to the functional relationship between the actual measurement and the motion state, the nonlinear measurement model can be constructed as
\begin{equation}
    \left[ d_{\mathrm{s}, n}, v_{\mathrm{s}, n}, \theta_{\mathrm{s}, n} \right] ^T
    = 
    \mathbf{h}( \boldsymbol{\xi }_{n} ) + \tilde{\boldsymbol{\omega}}_{n} , 
    \label{eq:measurement}
\end{equation}
where $\mathbf{h} \left( \cdot \right)$ stands for the nonlinear measurement function vector as
\vspace{-1mm}
\begin{displaymath}
    \mathbf{h}( \boldsymbol{\xi }_{n} ) \!=\!
    \left[ 
	\sqrt{x_{n}^{2} \!+\! y_{n}^{2}}, ~
	\left( \dot{x}_{n}x_{n} \!+\! \dot{y}_{n}y_{n} \right) /d_{n}, ~
	\mathrm{arc}\tan \left( y_{n}/x_{n} \right)
    \right]^T \!\!\!,
 \vspace{-1mm}
\end{displaymath}
and $\tilde{\boldsymbol{\omega}}_{n}$ denotes the measurement noise, which follows a zero-mean Gaussian random distribution with covariance
\begin{equation}
\begin{aligned}
    \mathbf{\Psi }_{n} 
    &= \mathrm{diag}\left( \sigma^2_{d,n} , \sigma^2_{v, n}  , \sigma^2_{\theta, n} \right).
\end{aligned}
\end{equation}
Here, $\sigma^2_{d,n} $,$ \sigma^2_{v, n}  $ and $ \sigma^2_{\theta, n}$ are inversely proportional to the SNR \cite{FLiu:20TWC, yuan2019scaled} and can be calculated as $\sigma^2_{d,n} = b_d \left(  \mathrm{SNR}_n B_{\mathrm{s}}^{2} \right) ^{-1}$, $\sigma^2_{v, n} = b_v \left( \mathrm{SNR}_n T_{\mathrm{e},n}^{2} \right) ^{-1}$ and $\sigma^2_{\theta, n} = b_{\theta} \left( \mathrm{SNR}_n / W_{\mathrm{null}} \right) ^{-1}$, where $B_{\mathrm{s}}^{2}$ and $T_{\mathrm{e}}$ represent the effective bandwidth and time duration for the sensing operation, respectively, and $W_{\mathrm{null}}$ indicates the null-to-null beam width of the receive antenna. The values of constants $b_d$, $b_v$ and $b_{\theta}$ depend on the specific system \cite{FLiu:20TWC}.

According to \cite{PCRB:98TSP,FDong:23TWC}, the PCRB characterizes the minimum achievable error covariance of any unbiased estimator, and is given by the inverse of the posterior information matrix. The PCRB is model-dependent rather than tied to a specific tracking filter. Within the Bayesian framework, the posterior (Fisher) information matrix $\mathbf{J}_{\boldsymbol{\xi},n}$ is computed as
\begin{equation}
\begin{aligned}
    \mathbf{J}_{\boldsymbol{\xi}, n} 
    % &= \mathbf{J}_{\mathrm{P},n}+ \mathbf{J}_{\mathrm{D},n} \\
    &= \left( \mathbf{\Phi}_n + \mathbf{F}_n  \mathbf{J}_{\boldsymbol{\xi }_{n-1}}^{-1}  \mathbf{F}_n^T \right) ^{-1}
    + \mathbf{H}_{n}^{T} \mathbf{\Psi}_{n}^{-1} \mathbf{H}_{n},
\end{aligned}
\label{eq:PCRB_J}
\end{equation}
where $\mathbf{H}_{n}$ represents the Jacobian matrix of the measurement function $\mathbf{h}( \boldsymbol{\xi }_{n} )$. The first term corresponds to the prior information, while the second term indicates the current measurement information. In the predicted PCRB, the latter is replaced by its pre-update counterpart.

Therefore, the PCRB can be expressed as
\begin{equation}
    \Gamma_n = \mathrm{tr} \left( \mathbf{J}_{\boldsymbol{\xi},n}^{-1} \right) .
    \label{eq:PCRB}
\end{equation}
The actual PCRB is evaluated after the measurement update at slot $n$, whereas the predicted PCRB is computed beforehand from the predicted state and prior information \cite{FDong:23TWC}. Thus, the former reflects the achievable post-update tracking accuracy, while the latter contains the pre-update state uncertainty. For notational clarity, for the remainder of this paper, $\Gamma_n$ denotes the actual PCRB at slot $n$, $\Gamma_{\hat{n}}$ defines the most recent available actual PCRB, and $\Gamma_n^{\mathrm{p}}$ refers to the predicted PCRB at slot $n$. The PCRB is model-dependent, and other recursive filters such as the UKF or particle filter can also be incorporated with consistent state updates.

\vspace{-2mm}
\section{SAC-based Adaptive Power Allocation for ISAC Tracking}
\label{sec:Proposed}
\vspace{-1mm}

In a continuous tracking scenario, the key challenge lies in adaptively allocating power to maintain reliable tracking accuracy while sustaining communication performance. In this section, we first formulate the objective problem and then introduce the proposed RL-based power allocation scheme for consecutive time slots.

\vspace{-2mm}
\subsection{Problem Description}
\vspace{-1mm}

Our objective is to exploit the prior information from the previous time slots to maximize the communication rate while maintaining reliable tracking accuracy, thereby avoiding power waste. As this work focuses on power-allocation designs rather than an estimator, the PCRB is used as a metric of the achievable tracking reliability. A smaller PCRB indicates a lower bound on the estimation error covariance under the adopted motion and measurement models. Compared with realized position or distance errors, which depend on a specific estimator and noise realization, the PCRB provides a control-oriented metric directly linked to the sensing quality and power allocation. The objective problem can be formulated as
\begin{align}
    (\mathbf{P1}) ~ 
    & \max_{ p_{\mathrm{c},n}, ~ p_{\mathrm{s},n} } ~ C_{n}
    \label{op:obj_func} \\
    \text{ s.t. }  
    & p_{\mathrm{c},n} > 0, p_{\mathrm{s},n} \geq 0, ~ \tag{\ref{op:obj_func}{a}} \label{const:P_LB}\\
    & p_{\mathrm{c},n} + p_{\mathrm{s},n} = P_{\max},  \tag{\ref{op:obj_func}{b}} \label{const:P_sum} \\
    & \Gamma_{n} \leq \overline{\Gamma} ,  ~\forall n, \tag{\ref{op:obj_func}{c}} \label{const:PCRB} 
    % &  \notag
\end{align}
where $\overline{\Gamma} $ denotes the maximum PCRB threshold. Constraint \eqref{const:PCRB} guarantees the reliability of tracking accuracy.

We may easily transform $(\mathbf{P1})$ into a convex form with the aid of fractional programming after converting \eqref{const:PCRB} into a convex constraint based on \cite{FDong:23TWC}, and the CVX may be employed. However, the actual PCRB is only available after obtaining the current measurement, while power allocation must be determined in advance. Therefore, \eqref{const:PCRB} cannot be directly used in model-based optimization. To make this implementable, one may consider employing the predicted PCRB instead of the actual PCRB, as in \cite{FDong:23TWC} and \cite{MLi:25WCL}, but it still suffers from the discrepancy between the predicted and actual PCRB. Moreover, since the PCRB includes an action-independent prior term, CVX-based optimization that considers only the current time slot may become infeasible when the PCRB stays above the threshold $\overline{\Gamma}$. Therefore, we address this challenge through an RL approach that leverages past experiences to prevent infeasible cases and dynamically adjust power allocation. In particular, our RL design uses the most recent available actual PCRB, the current predicted PCRB, and the sensing interval to evaluate the tracking status over time.

\vspace{-2mm}
\subsection{SAC-based Power Allocation}
\vspace{-1mm}

To solve $(\mathbf{P1})$, we deploy an RL agent at the ISAC BS to adaptively allocate power. The BS agent operates under partial observations on the sensing side. First, the true target state is not directly available, and thus prior RL-based studies use the predicted state as the observation \cite{YWang:24TWC}. Also, the agent may skip sensing when recent estimation satisfies a reliability criterion, which causes larger prediction uncertainty. Our scheme solves these issues by monitoring estimate reliability and performing sensing when only needed.

\textit{1) Observation:}
To handle partial observation, we include the PCRB in the state to quantify the reliability of the available target state estimation. The state vector $\mathbf{s}_n$ is constructed as 
\begin{equation}
    \mathbf{s}_n = \left[ \Gamma_{\hat{n}}, p_{\mathrm{s},\hat{n}}, \Gamma^{\mathrm{p}}_n, \bar{p}, g_n, | \mathbf{h}_{\mathrm{c}}^{H} \mathbf{u}|^2, n \right]
\end{equation}
where $\Gamma_{\hat{n}}$ and $p_{\mathrm{s},\hat{n}}$ represent the actual PCRB and the sensing power at the most recent slot $\hat{n}$ with ISAC mode, respectively, and $\Gamma^{\mathrm{p}}_n$ indicates the predicted PCRB with average power allocation $\bar{p} = P_{\max} /2$ at the current slot. At the beginning of each episode, these variables are initialized with a full-power beam-sweeping stage assuming that the initial target state is known.

\textit{2) Action:}
We define the action as the power allocation ratios as
\vspace{-1mm}
\begin{equation}
    \mathbf{a}_n = \left[ \frac{p_{\mathrm{c},n}}{P_{\max}} , \frac{p_{\mathrm{s},n}}{P_{\max}} \right].
    \label{eq:action}
    \vspace{-1mm}
\end{equation}
When interacting with the environment, the actual transmit power vector is given by $P_{\max} \mathbf{a}_n$. Since the available power budget is active in the considered allocation problem, the ratio-based action is equivalent to the absolute power allocation but provides a normalized representation of the simplex-constrained action space. This normalized form is convenient for policy learning and is also well-suited to the Dirichlet policy introduced later, which naturally generates continuous actions over the simplex. A detailed design of the policy that produces such ratio-based actions will be introduced later.

\textit{3) Reward:}
The reward function is designed to encourage efficient power allocation while maintaining a balance between sensing and communication. The power constraints are enforced through the policy network design, while the reward mainly addresses the PCRB constraint (12c) and the communication objective. When the PCRB constraint is satisfied, a rate-related reward is given. Otherwise, a PCRB-based penalty is imposed. A sensing interval-related penalty is also introduced to discourage excessive sensing skips, since long sensing gaps may increase prediction uncertainty and the risk of PCRB violation. In addition, an exponential form is adopted for the communication-related reward to bound its value and smooth the effect of $C_n$, thereby improving training stability. When most of the available power is assigned to communication, $C_n$ may become highly sensitive to the power-allocation decision, which can lead to large reward fluctuations during training. The exponential transformation is thus adopted to stabilize the learning process.

As a result, the reward can be calculated as 
\begin{equation}
    r_n \! = \!
    \begin{cases}
        \omega_{\mathrm{c}} \left( 1 - \exp(-0.1 C_n) \right) - \omega_{\mathrm{T}} g_n \!\!\!\!\!\! &, \text{if $\Gamma_{\hat{n}} \leqslant \overline{\Gamma}$}, \\
        \omega_{\mathrm{s}} \left( \overline{\Gamma} - \Gamma_{\hat{n}} \right) - \omega_T g_n \!\!\!\!\!\! &, \text{otherwise},\\
      \end{cases}
\end{equation}
where $\omega_{\mathrm{c}}$, $\omega_{\mathrm{T}}$, and $\omega_{\mathrm{s}}$ denote the weighting factors associated with the communication rate, the sensing-skip penalty, and the PCRB constraint, respectively. These weighting factors are chosen so that the rate-related reward and the penalty terms have comparable numerical scales during training. If the agent takes action with the ISAC mode at the current slot, $\Gamma_{\hat{n}}$ should be updated with the newest result when computing the reward.

With the proposed design, the agent can learn the relationship between sensing and communication performance, tracking history, and power allocation. Here, the PCRB captures the effect of sensing intervals on the measurement accuracy, but does not acquire the stochastic nature of target motion. We thus model the environment as the Markov decision process (MDP) in which the next state is drawn from a transition distribution conditioned on the current state and action. Therefore, we adopt the soft actor-critic (SAC) algorithm with double Q-networks, which enables stochastic policy learning in environments with uncertain dynamics and long-term objectives~\cite{haarnoja2018soft_v2}. In addition, the stochastic exploration promoted by SAC is beneficial for mitigating temporary tracking-reliability violations during learning, since it allows the agent to exploit past experience while retaining sufficient exploration capability.

As the power allocation is constrained by a simplex form, we employ the Dirichlet-based stochastic policy \cite{tian2022prescriptive} as
\vspace{-1mm}
\begin{equation}
    \pi_\phi (\mathbf{a}_n | \mathbf{s}_n) = \frac{1}{ B(z_\phi (\mathbf{s}_n)) } \prod_{i=1}^{N} \mathbf{a}_i^{z_i (\mathbf{s}_n) -1} ,
\vspace{-1mm}
\end{equation}
where $\phi$ denotes the parameters of the policy network, which outputs the Dirichlet distribution parameters $z$, and $B(z)$ indicates the multivariate beta function. This policy formulation ensures that the actions lie within valid power allocation space without requiring any additional transformation and Jacobian corrections (e.g., softmax or tanh), and its policy gradient estimator is provably unbiased under the reparameterization trick \cite{tian2022prescriptive}. 

Two critic networks are trained independently to minimize the soft Bellman residual loss \cite{haarnoja2018soft_v2} as
\vspace{-1mm}
\begin{equation}
    J_Q \! \left( \theta_i \right) \! = \! 
    \mathbb{E}_{(\! \mathbf{s}_n \!,\! \mathbf{a}_n \!) \sim \mathcal{D}} \! \left[ \frac{1}{2} \left( Q_{\theta_i} \! (\mathbf{s}_n, \mathbf{a}_n) \!-\! \hat{Q}(\mathbf{s}_n, \mathbf{a}_n)  \right)^2 \right] \! , 
\end{equation}
where $\mathcal{D}$ stands for the experience replay buffer, $Q_{\theta_i}(\cdot)$ denotes the parametrized critic networks for $i \in \{1,2\}$, and $\hat{Q}(\cdot)$ equals the target Q-value as
\begin{align}
    \hat{Q}(  \mathbf{s}_n, \mathbf{a}_n  ) 
     =
    & r_n \!+\! \gamma \mathbb{E}_{\mathbf{a}_{n+1} \sim \pi_{\phi} (\cdot|\mathbf{s}_{n+1}) } \! \Big[ \min_{i=1,2}  Q_{\bar{\theta}_i} (\mathbf{s}_{n+1}, \mathbf{a}_{n+1}) \notag\\
    & - \beta \log \! \pi_\phi (\mathbf{a}_{n+1} | \mathbf{s}_{n+1}) \Big].
\end{align}
Here, $\gamma$ represents the discount factor, $Q_{\bar{\theta}_i}(\cdot)$ denotes the target critic networks, and $\beta$ is a positive temperature factor to encourage the exploration of actions.

Then, the actor network is trained to minimize the entropy-regularized policy loss as
\begin{equation}
    J_{\pi} \! \left( \! \phi \! \right) 
    \!=\!
    \mathbb{E}_{\mathbf{s}_n \! \sim \! \mathcal{D},  \mathbf{a}_n \! \sim \! \pi_\phi} \!\!\! \left[ \! \beta \! \log( \pi_{\phi}( \mathbf{a}_n \! | \mathbf{s}_n ) ) \!-\!\!  \min_{i=1,2} \! Q_{\theta_i} \! (\mathbf{s}_n, \! \mathbf{a}_n) \right] \!\! .\!\!\!
\end{equation}
To balance exploration and exploitation, the entropy coefficient $\beta$ is automatically tuned during training to maintain the policy entropy near a target value \cite{haarnoja2018soft_v2}.

\vspace*{-3mm}
\section{Numerical Results and Conclusion} 
\label{sec:NumericalResults}
\vspace*{-1mm}

\begin{table}
    \centering
    \caption{Simulation Setup}
    \vspace*{-2mm}
    \begin{tabular}[ht!]{|l|c|}
        \hline
        Number of BS antennas $M$  &  $16$ \\
        \hline
        Total number of time slots $N$  &  $80$ \\ \cline{1-2}
        % \hline
        % Length of time slot $T_{\mathrm{s}}$  &  $0.01$ s  \\ \cline{1-2}
        \hline
        Carrier frequency  &  30 GHz \\ \cline{1-2}
        \hline
        Maximum transmit power at BS $P_{\max}$  &  $40$ dBm \\ \cline{1-2}
        \hline
        Path-loss exponent $\eta$  &  $2$ \\ \cline{1-2}
        \hline
        Noise variance of communication $\sigma_{\mathrm{c}}^2$  &  $-93$ dBm \\ \cline{1-2}
        \hline
        Noise variance of sensing $\sigma_{\mathrm{s}}^2$  &  $0$ dBm \\ \cline{1-2}
        \hline
        % Noise variance of RCS $\sigma_{\alpha_n}^2$  &  $\left[0,\frac{\pi}{36}\right]$ \\ \cline{1-2}
        % \hline
        Process noise intensity $q_{\mathrm{s}}$  &  $0.4$ m²/s³ \\ \cline{1-2}
        \hline
        Efficient bandwidth for sensing $B_{\mathrm{s}}$  &  $50$ MHz \\ \cline{1-2}
        \hline
        Mini-batch size $|\mathcal{B}|$  &  $1024$  \\ \cline{1-2}
        \hline
        Reward discount factor $\gamma$  &  0.99  \\ \cline{1-2}
        \hline
        Learning rate of actor network  &  $10^{-4}$  \\ \cline{1-2}
        \hline
        Learning rate of critic network  &  $3 \times 10^{-4}$  \\ \cline{1-2}
        \hline
        Learning rate of entropy temperature  &  $10^{-5}$  \\ \cline{1-2}
        \hline
        Initial entropy coefficient  &  0.3  \\ \cline{1-2}
        \hline
        Target entropy  &  -2  \\ \cline{1-2}
        \hline
    \end{tabular}
    \label{simul_SetUp}
    \vspace*{-4mm}
\end{table}

In this section, we evaluate the performance of the proposed SAC-based power allocation scheme with the Dirichlet policy, and compare with the following baselines: 1) \textit{SAC with predicted state:} the sensing-related state is replaced by the predicted target state at the current slot, 2) \textit{CVX-based scheme:} the BS allocates all power to sensing when an infeasible case happens with the CVX solver. The simulation is conducted in a two-dimensional coordinate system with the BS located at $(0 \mathrm{m}, ~ 0 \mathrm{m})$. The communication user is placed at $(0 \mathrm{m}, ~ 80 \mathrm{m})$, while the sensing target is located at $45^{\circ}$, $(40\sqrt{2} \mathrm{m},~ 40\sqrt{2} \mathrm{m})$. The target moves with a $5$ m/s velocity towards the positive x-axis direction, and its motion follows the evolution model with process noise as introduced in Sec. \RNum{3}. The constants in the measurement noise variance expression \cite{FLiu:20TWC} are set to $b_d=1$, $b_v = 6.7 \times 10^{-7}$ and $b_{\theta}= 2 \times 10^4$. The length of a time slot is set as $0.01$ s. The PCRB threshold is $\overline{\Gamma}=0.02$. Further simulation parameters are provided in Table \ref{simul_SetUp}.

\begin{figure}[bp]
\vspace*{-4mm}
    \begin{center}
    \includegraphics[width=2.6in, trim=90 260 110 280, clip]{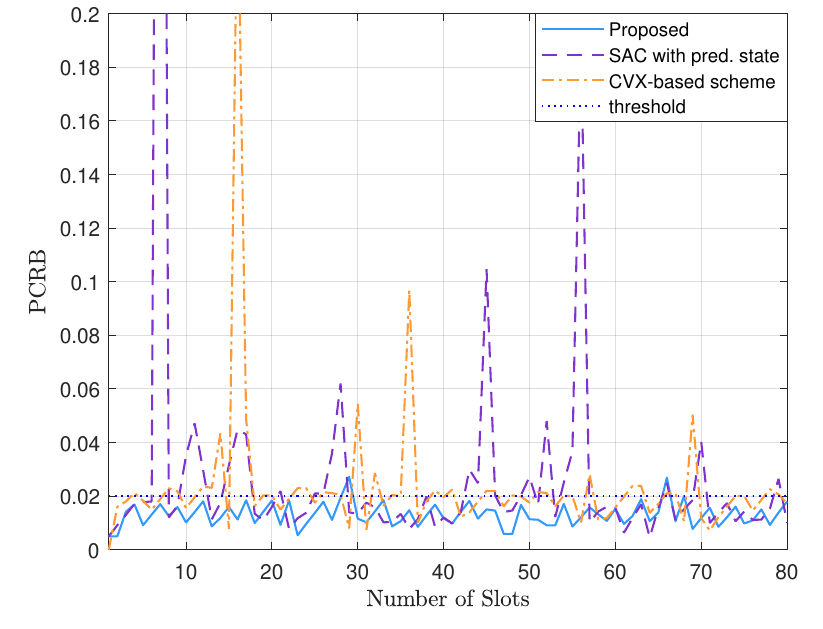}  % 2col
    \end{center}
    \vspace{-4mm}
    \caption{Feasibility of PCRB constraints across different schemes}
    \label{fig:FeasibilityCheck}
    \vspace*{-6mm}
\end{figure}

\begin{figure*}[htbp]
\centering
	\subfloat[Total reward]{
    \includegraphics[width = 0.3\textwidth, trim=90 260 110 280, clip]{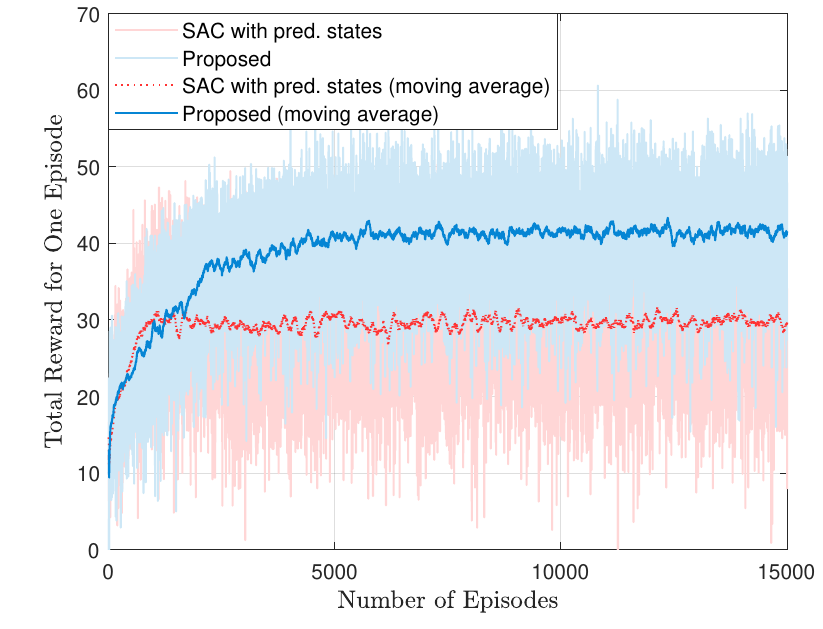}  % 2col
    \label{fig:results_reward}}
	\subfloat[Sensing PCRB]{
    \includegraphics[width = 0.3\textwidth, trim=90 260 110 280, clip]{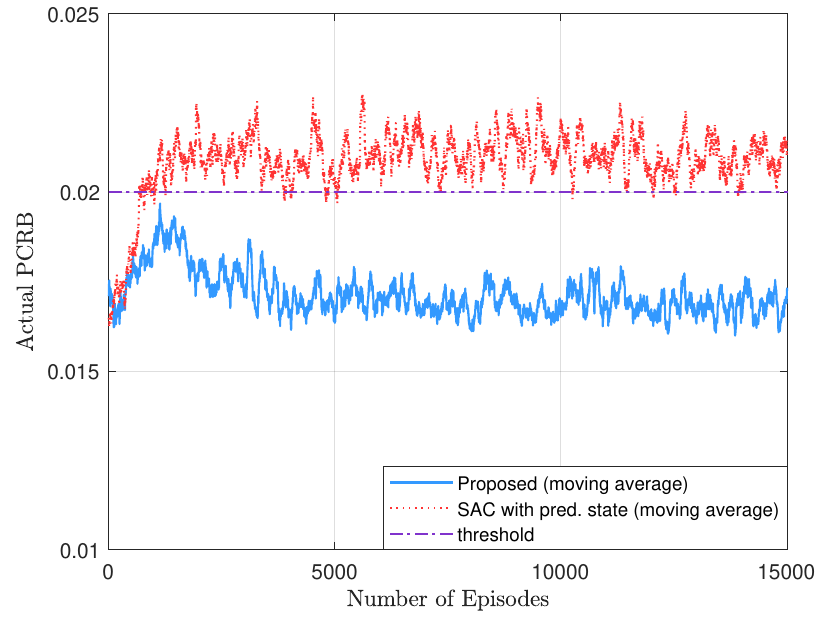}  % 2col
    \label{fig:results_PCRB}}
	\subfloat[Communication rate]{
    \includegraphics[width = 0.3\textwidth, trim=90 260 110 280, clip]{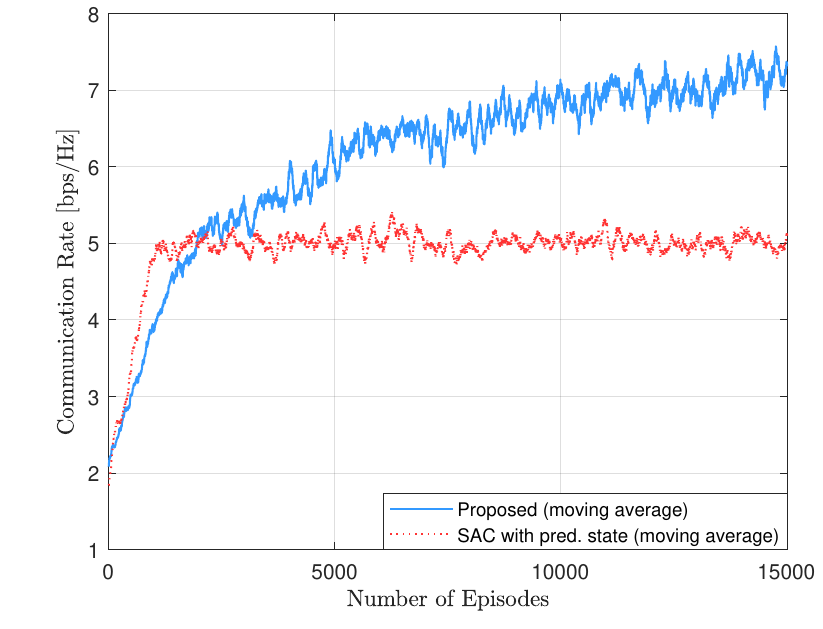}  % 2col
    \label{fig:results_rate}} 
\caption{Convergence Behavior}
\label{fig:ConvergenceBehavior}
\vspace*{-5mm}
\end{figure*}

Fig. \ref{fig:FeasibilityCheck} presents a single test-episode result using the trained policies for feasibility checking. The proposed scheme consistently maintains the PCRB below the threshold for almost the entire episode, whereas other approaches occasionally fail to satisfy the requirement. The SAC with the predicted state exhibits sporadic peaks, reflecting its limited adaptability due to prediction errors. Also, the CVX-based scheme incurs 40\% infeasible slots, highlighting that the CVX cannot be applied in our problem.

Fig. \ref{fig:ConvergenceBehavior} summarizes the training convergence of two RL-based schemes. Fig. \ref{fig:ConvergenceBehavior}\subref{fig:results_reward} illustrates the training rewards of different schemes. We can see that the proposed method converges after 6000 episodes with higher rewards. The proposed scheme yields smoother reward curves, while the other scheme suffers from larger fluctuations caused by prediction errors. This indicates that leveraging the PCRB information results in more effective power allocation between sensing and communication. Fig. \ref{fig:ConvergenceBehavior}\subref{fig:results_PCRB} and Fig. \ref{fig:ConvergenceBehavior}\subref{fig:results_rate} further confirm sensing and communication performance, respectively. The SAC with the predicted state scheme fails to keep the PCRB within the required threshold and does not improve communication performance during training. The inaccuracy in the predicted states prevents the agent from adaptively balancing the trade-off between sensing and communication. In contrast, the proposed scheme progressively learns to enhance communication performance while maintaining sensing accuracy, demonstrating its ability to adaptively allocate power in mobility scenarios.

In conclusion, the simulation results verify that the proposed scheme effectively balances sensing and communication. The proposed scheme avoids unnecessary sensing while meeting PCRB requirements. Our results offer new insights for resource-efficient ISAC and motivate extensions to more complex scenarios. An interesting direction for future work is to extend the proposed framework to multi-user and multi-target ISAC scenarios. The corresponding reward design and resource-coupling management become significantly more challenging. Furthermore, the framework may be extended to longer decision intervals, where target-state variations may become more significant and require corresponding changes in the motion model, measurement model, and tracking-quality evaluation. Future work may also consider alternative tracking filters to further examine the robustness of the proposed framework.

\bibliographystyle{ieeetr}
\begingroup
\renewcommand{\baselinestretch}{0.94}
\bibliography{AZREF}

@ARTICLE{FDong:23TWC,
  author={Dong, Fuwang and Liu, Fan and Cui, Yuanhao and Wang, Wei and Han, Kaifeng and Wang, Zhiqin},
  journal={IEEE Trans. Wirel. Commun.}, 
  title={{Sensing as a Service in 6G Perceptive Networks: A Unified Framework for ISAC Resource Allocation}}, 
  year={2023},
  month={Nov.},
  volume={22},
  number={5},
  pages={3522-3536},
  doi={10.1109/TWC.2022.3219463}}

@ARTICLE{PCRB:98TSP,
  author={Tichavsky, P. and Muravchik, C.H. and Nehorai, A.},
  journal={IEEE Trans. Signal Process}, 
  title={{Posterior Cramer-Rao Bounds for Discrete-time Nonlinear Filtering}}, 
  year={1998},
  month={May},
  volume={46},
  number={5},
  pages={1386-1396},
  doi={10.1109/78.668800}}

@ARTICLE{FLiu:20TWC,
  author={Liu, Fan and Yuan, Weijie and Masouros, Christos and Yuan, Jinhong},
  journal={IEEE Trans Wirel. Commun.}, 
  title={{Radar-Assisted Predictive Beamforming for Vehicular Links: Communication Served by Sensing}}, 
  year={2020},
  volume={19},
  number={11},
  pages={7704-7719},
  doi={10.1109/TWC.2020.3015735}}

@ARTICLE{yuan2019scaled,
  title={{Scaled Accuracy based Power Allocation for Multi-target Tracking with Colocated MIMO Radars}},
  author={Yuan, Ye and Yi, Wei and Kirubarajan, Thia and Kong, Lingjiang},
  journal={Signal Processing},
  volume={158},
  pages={227--240},
  year={2019},
  month={Mar.},
  publisher={Elsevier}
}

@ARTICLE{YWang:24TWC,
  author={Wang, Yijie and Lou, Mengting and Qian, Wanyun and Bai, Yechao and Tang, Lan and Liang, Ying-Chang},
  journal={IEEE Trans. Veh. Technol.}, 
  title={{Intelligent Beam Tracking in Radar-Assisted MIMO-OFDM Communication Systems}}, 
  year={2024},
  month={June},
  volume={73},
  number={11},
  pages={16774-16789},
  doi={10.1109/TVT.2024.3419000}}

@ARTICLE{MLi:25WCL,
  author={Li, Ming and Dong, Fuwang and Liu, Tao and Liu, Fan},
  journal={IEEE Wirel. Commun. Lett.}, 
  title={{EKF-Based Beamforming Design for Joint Beam Tracking and Communication Systems}}, 
  month={Early Access,},
  year={2025},
  volume={},
  number={},
  pages={},
  doi={10.1109/LWC.2025.3583503}}

@ARTICLE{KAbboud:16,
  author={Abboud, Khadige and Omar, Hassan Aboubakr and Zhuang, Weihua},
  journal={IEEE Trans. Veh. Technol.}, 
  title={{Interworking of DSRC and Cellular Network Technologies for V2X Communications: A Survey}}, 
  year={2016},
  month={Dec.},
  volume={65},
  number={12},
  pages={9457-9470},
  doi={10.1109/TVT.2016.2591558}}

@ARTICLE{MRahman:20TAES,
  author={Rahman, Md. Lushanur and Zhang, J. Andrew and Huang, Xiaojing and Guo, Y. Jay and Heath, Robert W.},
  journal={IEEE Trans. Aerosp. Electron. Syst.}, 
  title={{Framework for a Perceptive Mobile Network Using Joint Communication and Radar Sensing}}, 
  year={2020},
  month={June},
  volume={56},
  number={3},
  pages={1926-1941},
  doi={10.1109/TAES.2019.2939611}}

@ARTICLE{CBarneto:22TCom,
  author={Barneto, Carlos Baquero and Riihonen, Taneli and Liyanaarachchi, Sahan Damith and Heino, Mikko and González-Prelcic, Nuria and Valkama, Mikko},
  journal={IEEE Trans. Commun.}, 
  title={{Beamformer Design and Optimization for Joint Communication and Full-Duplex Sensing at mm-Waves}}, 
  year={2022},
  month={Dec.},
  volume={70},
  number={12},
  pages={8298-8312},
  doi={10.1109/TCOMM.2022.3218623}}

@ARTICLE{JLee:22TVT,
  author={Lee, Joash and Cheng, Yanyu and Niyato, Dusit and Guan, Yong Liang and González, David G.},
  journal={IEEE Trans. Veh. Technol.}, 
  title={{Intelligent Resource Allocation in Joint Radar-Communication With Graph Neural Networks}}, 
  year={2022},
  month={Oct.},
  volume={71},
  number={10},
  pages={11120-11135},
  doi={10.1109/TVT.2022.3187377}}

@ARTICLE{JZhang:22,
  author={Zhang, J. Andrew and Rahman, Md. Lushanur and Wu, Kai and Huang, Xiaojing and Guo, Y. Jay and Chen, Shanzhi and Yuan, Jinhong},
  journal={IEEE Commun. Surv. Tutor.}, 
  title={{Enabling Joint Communication and Radar Sensing in Mobile Networks—A Survey}}, 
  year={2022},
  month={First quarter},
  volume={24},
  number={1},
  pages={306-345},
  doi={10.1109/COMST.2021.3122519}}

@ARTICLE{KMeng:23TWC,
  author={Meng, Kaitao and Wu, Qingqing and Ma, Shaodan and Chen, Wen and Wang, Kunlun and Li, Jun},
  journal={IEEE Trans. Wirel. Commun.}, 
  title={{Throughput Maximization for UAV-Enabled Integrated Periodic Sensing and Communication}}, 
  year={2023},
  month={Jan.},
  volume={22},
  number={1},
  pages={671-687},
  doi={10.1109/TWC.2022.3197623}}

@ARTICLE{ZXiao:25TWC,
  author={Xiao, Zhiqiang and Zeng, Yong and Wen, Fuxi and Zhang, Zaichen and Ng, Derrick Wing Kwan},
  journal={IEEE Trans. Wirel. Commun.},
  title={{Integrated Sensing and Channel Estimation by Exploiting Dual Timescales for Delay-Doppler Alignment Modulation}}, 
  year={2025},
  month={Jan.},
  volume={24},
  number={1},
  pages={415-429},
  doi={10.1109/TWC.2024.3493255}}

@ARTICLE{PRaviteja:18TWC,
  author={Raviteja, P. and Phan, Khoa T. and Hong, Yi and Viterbo, Emanuele},
  journal={IEEE Trans. Wirel. Commun.}, 
  title={{Interference Cancellation and Iterative Detection for Orthogonal Time Frequency Space Modulation}}, 
  year={2018},
  month={Aug.},
  volume={17},
  number={10},
  pages={6501-6515},
  doi={10.1109/TWC.2018.2860011}}

@ARTICLE{NGP:24Proc,
  author={González-Prelcic, Nuria and Furkan Keskin, Musa and Kaltiokallio, Ossi and Valkama, Mikko and Dardari, Davide and Shen, Xiao and Shen, Yuan and Bayraktar, Murat and Wymeersch, Henk},
  journal={Proc. IEEE.}, 
  title={{The Integrated Sensing and Communication Revolution for 6G: Vision, Techniques, and Applications}}, 
  year={2024},
  month={May},
  volume={112},
  number={7},
  pages={676-723},
  doi={10.1109/JPROC.2024.3397609}}

@ARTICLE{PKumari:18TVT,
  author={Kumari, Preeti and Choi, Junil and González-Prelcic, Nuria and Heath, Robert W.},
  journal={IEEE Trans. Veh. Technol.}, 
  title={{IEEE 802.11ad-Based Radar: An Approach to Joint Vehicular Communication-Radar System}}, 
  year={2018},
  month={Nov.},
  volume={67},
  number={4},
  pages={3012-3027},
  doi={10.1109/TVT.2017.2774762}}

@ARTICLE{JLi:24IOTJ,
  author={Li, Jiapeng and Shao, Xiaodan and Chen, Feng and Wan, Shaohua and Liu, Chang and Wei, Zhiqiang and Wing Kwan Ng, Derrick},
  journal={IEEE Internet Things J.}, 
  title={{Networked Integrated Sensing and Communications for 6G Wireless Systems}}, 
  year={2024},
  month={May},
  volume={11},
  number={17},
  pages={29062-29075},
  doi={10.1109/JIOT.2024.3406598}}

@ARTICLE{AGraff:23TVT,
  author={Graff, Andrew and Chen, Yun and González-Prelcic, Nuria and Shimizu, Takayuki},
  journal={IEEE Trans. Veh. Technol.}, 
  title={{Deep Learning-Based Link Configuration for Radar-Aided Multiuser mmWave Vehicle-to-Infrastructure Communication}}, 
  year={2023},
  month={Jan},
  volume={72},
  number={6},
  pages={7454-7468},
  doi={10.1109/TVT.2023.3239227}}

@article{tian2022prescriptive,
  title={{A prescriptive Dirichlet power allocation policy with deep reinforcement learning}},
  author={Tian, Yuan and Han, Minghao and Kulkarni, Chetan and Fink, Olga},
  journal={{Reliability Engineering \& System Safety}},
  volume={224},
  pages={108529},
  year={2022},
  month={Aug.},
  publisher={Elsevier}
}

@article{haarnoja2018soft_v2,
  title={Soft actor-critic algorithms and applications},
  author={Haarnoja, Tuomas and Zhou, Aurick and Hartikainen, Kristian and Tucker, George and Ha, Sehoon and Tan, Jie and Kumar, Vikash and Zhu, Henry and Gupta, Abhishek and Abbeel, Pieter and others},
  journal={arXiv preprint arXiv:1812.05905},
  year={2018}
}

@ARTICLE{ZPu:24TGCN,
  author={Pu, Zhiwei and Wang, Wei and Lao, Zhiwei and Yan, Ye and Qin, Hongde},
  journal={IEEE Trans. Green Commun. Netw.}, 
  title={Power Allocation of Integrated Sensing and Communication System for the Internet of Vehicles}, 
  year={2024},
  month={Dec.},
  volume={8},
  number={4},
  pages={1717-1728},
  doi={10.1109/TGCN.2024.3391015}}

@ARTICLE{LabMintae:24IOT,
  author={Kim, Mintae and Lee, Hoon and Hwang, Sangwon and Debbah, Mérouane and Lee, Inkyu},
  journal={IEEE Internet Things J.}, 
  title={{Cooperative Multiagent Deep Reinforcement Learning Methods for UAV-Aided Mobile Edge Computing Networks}}, 
  year={2024},
  month={Dec.},
  volume={11},
  number={23},
  pages={38040-38053},
  doi={10.1109/JIOT.2024.3447090}}

@ARTICLE{LabSangwon:25IOT,
  author={Hwang, Sangwon and Lee, Hoon and Kim, Mintae and Lee, Inkyu},
  journal={IEEE Internet Things J.}, 
  title={{Multiagent Deep Reinforcement Learning for Decentralized Multi-UAV Mobile Edge Computing Networks}}, 
  year={2025},
  month={May},
  volume={12},
  number={10},
  pages={14484-14497},
  doi={10.1109/JIOT.2025.3527016}}

@ARTICLE{LabSubin:21TVT,
  author={Eom, Subin and Lee, Hoon and Park, Junhee and Lee, Inkyu},
  journal={IEEE Trans. Veh. Technol.}, 
  title={{Asynchronous Protocol Designs for Energy Efficient Mobile Edge Computing Systems}}, 
  year={2021},
  month={Jan.},
  volume={70},
  number={1},
  pages={1013-1018},
  doi={10.1109/TVT.2020.3044073}}

@ARTICLE{SlotLength5G,
  author={Chakrapani, Arvind},
  journal={IEEE Access}, 
  title={{On the Design Details of SS/PBCH, Signal Generation and PRACH in 5G-NR}}, 
  year={2020},
  month={July},
  volume={8},
  number={},
  pages={136617-136637},
  doi={10.1109/ACCESS.2020.3010500}}
\endgroup
\end{document}